\documentclass[twocolumn,showpacs,fleqn,nofootinbib]{revtex4}
\usepackage{amsmath}
\usepackage{graphicx}
\usepackage{float}
\usepackage{subfigure}
\usepackage{verbatim}
\usepackage{color}
\usepackage{comment}
\usepackage{hyperref}
\usepackage{epstopdf}
\usepackage{booktabs}
\usepackage[utf8]{inputenc}

\newcommand\tr{\mathop{\rm Tr}\nolimits}
\renewcommand\Im{\mathop{\rm Im}\nolimits}

\newcommand\vk{{\vec k}}
\newcommand\vl{{\vec l\,}}
\newcommand\vp{{\vec p\,}}
\newcommand\vq{{\vec q\,}}
\newcommand{\pom}{I\!\!P}                
\def\lsim{\mathrel{\rlap{\lower4pt\hbox{\hskip1pt$\sim$}}
    \raise1pt\hbox{$<$}}}                
\def\gsim{\mathrel{\rlap{\lower4pt\hbox{\hskip1pt$\sim$}}
    \raise1pt\hbox{$>$}}}                

\begin{document}
	
	\title{Determination of the entanglement entropy in elastic scattering using model-independent method for hadron femtoscopy}
	\author{G.S. Ramos and M.V.T. Machado}
	
	\affiliation{High Energy Physics Phenomenology Group, GFPAE  IF-UFRGS \\
		Caixa Postal 15051, CEP 91501-970, Porto Alegre, RS, Brazil}

\begin{abstract}
The entanglement entropy of two-body elastic scattering at high energies is studied by using the model-independent L\'evy imaging method for investigating the hadron structure. It is considered the finite entropy in the momentum Hilbert space properly regularized and the results are compared to recent evaluation using the diffraction peak approximation. We present the entropy for RHIC, Tevatron and LHC energies pointing out the underlying uncertainties.
\end{abstract}	

\maketitle
	
\section{Introduction}	

Entanglement entropy is a current hot topic in high energy physics, quantum gravity and quantum field theory (see, for instance the reviews \cite{Berges:2020fwq,Headrick:2019eth,Rangamani:2016dms}), measuring the depart from a pure quantum state by a particle system. Recently, the confinement of partons inside hadrons has been investigated as an example of quantum entanglement due to their correlations and where only a part of the Hilbert space of quark-gluon system is probed by the scattering probe (projectile electrons, virtual photons, proton, etc).  It has been demonstrated that in the deep inelastic scattering of leptons off hadrons (DIS) a non-zero von Neumann entropy is obtained from different configurations of quasi-free incoherent partons inside probed hadron because of their quantum entanglement \cite{Kharzeev:2017qzs}. Specifically focusing on gluons the entropy associated with their production  \cite{Kutak:2011rb} is obtained taking into account perturbative QCD saturation formalism. An upper bound on the entropy of gluons was found, which  is deeply connected with the measured hadron multiplicity in proton-proton collisions at high energies.  Along similar lines,  the entropy of quarks and gluons using  the semi classical counterpart of von Neumann, i.e. the  Wehrl entropy, has been derived in Ref. \cite{Hagiwara:2017uaz}. The  phase space  QCD Wigner and Husimi distributions for partons are taken into account and Wehrl entropy is  given  in terms of the gauge invariant matrix element of the parton field operators.  Moreover, within the color glass condensate (CGC) formalism for the fast hadron wavefunction the entropy of soft gluons was obtained in \cite{Kovner:2015hga} and the evolution equations for the effective CGC density matrix has been investigated \cite{Armesto:2019mna,Li:2020bys}. In the same context, the entropy associated to a partial set of measurements on a quantum state  named as  entropy of ignorance \cite{Duan:2020jkz} was defined. It  is equal to the Boltzmann entropy of a classical system of quarks and gluons and is similar to entanglement entropy at high momenta.

Recently, we computed the entanglement entropy of gluons within the nucleons and nuclei by considering analytical parameterizations for the gluon distribution function (PDF) in the context of QCD saturation formalism \cite{Ramos:2020kaj}. It was compared to  current extractions of entropy using hadron multiplicities in DIS and proton-proton collisions at the LHC \cite{Tu:2019ouv}. The relation with other approaches for parton entropy as the CGC formalism and Wehrl entropy was investigated and the nuclear entanglement entropy per nucleon was addressed as well. Summarizing ideas,  the entanglement entropy, $S_{EE}=\ln[xg(Y,Q^2)]$, is determined by the gluon distribution, $xg(x,\mu^2)$, evaluated at a probing scale $\mu^2=Q^2$ at a rapidity $Y=\ln (1/x)$ ($x$ is the usual Bjorken variable). There is the identification of the gluon distribution with the average number of particles, $N$, such that $S_{EE}=\ln(N)$ in small-$x$ DIS.  In the large $Y$ limit the entanglement entropy is maximal meaning that the equipartition of micro-states maximizing them corresponds to the gluon saturation.  The extracted values from $ep$ DIS at  DESY-HERA and proton-proton collisions at the Large Hadron Collider (LHC)  are of order $S_{EE}\sim 2-3$ for $Y\simeq 7-9$  \cite{Ramos:2020kaj,Tu:2019ouv}, which is consistent with the entanglement entropy of the initial state partons derived within the nonlinear QCD evolution formalism \cite{Kharzeev:2017qzs}. These results are corroborated by recent determination \cite{Gotsman:2020bjc} that the multiplicity distribution of hadrons described by QCD evolution equations scales on $N$ in the form $\sigma_n/\sigma_{inel}=\frac{1}{N}(\frac{N-1}{N})^{n-1}$ and $S=\ln (N)$ corresponding to the high energy partonic state being maximally entangled. Here, the quantity  $\sigma_n$ is the cross section for producing $n$ hadrons  and with $\sigma_{inel}$ being the inelastic cross section. In studies presented in \cite{Ramos:2020kaj} the key quantity is the saturation scale, $Q_s(x)^2=Q_0^2(x/x_0)^{-\lambda}\sim e^{\lambda Y}$ ($\lambda\simeq 0.25$, $x_0\simeq 10^{-5}$ and $Q_0=1$ GeV), which is the typical gluon transverse momentum at very small $x$ or quite large rapidities $Y$. It was demonstrated that both the Werhl entropy and entanglement entropy from the CGC formalism  behave as $S_{EE},\,S_W\sim S_{\perp}Q_s^2$ with $S_{\perp}$ being the target transverse area. On the other hand, in the QCD dipole cascade formalism \cite{Kharzeev:2017qzs,Gotsman:2020bjc} it behaves asymptotically as $S_{EE}\sim Y\ln(Q_s^2/\mu^2)\sim Y^2$. The nuclear entanglement entropy was also investigated and the main result is that the nuclear Wehrl entropy behaves like $S_W\sim S_{\perp}^AQ_{s,A}^2\sim Ae^{\lambda Y}$ with the nuclear saturation scale being $Q_{s,A}^2\sim A^{1/3}Q_s^2$ and the nucleus transverse area given by $S_{\perp}^A=\pi R_A^2\sim A^{2/3}$. Therefore, the nuclear entanglement entropy for gluons inside nuclei is additive respect to the hadron one and consistent with $S_{EE}$ being an extensive variable.  Although it is tempting to follow the same parton saturation frameworks to treat the entropy of produced hadron in soft region in collider energies, here we will use a non-perturbative approach. The saturation scale provide us with a semihard scale at high energies, which allows to extend the perturbative analysis deep in the soft region as we shown in \cite{Peccini:2020ybx}. However, the measured observables as the differential elastic cross section at small$-t$ and total cross section should be dominated by non-perturbative aspects of QCD and a $S$-matrix approach is more appropriated.   

In this work we focus on the entanglement entropy generated by the two-body elastic scattering in the high energy limit. Now the underlying dynamics is given by non-perturbative sector of the QCD or in the Regge phenomenology by the soft Pomeron physics. In particular, we investigate the hadron-hadron strong interaction scattering including both the elastic ($A+B\rightarrow A+B$) and inelastic ($A+B\rightarrow X$) channels by using the S-matrix formalism. In the latter, the full Hilbert space of states is factorized into the Hilbert spaces of the initial and final states. We follow closely the Refs. \cite{Peschanski:2016hgk,Peschanski:2019yah}, where the reduced matrix, $\hat{\rho}_A$,  of the final state with two outgoing particles in an elastic scattering is computed in terms of the partial wave expansion of the two-body states. The entanglement entropy is obtained through the R\'enyi entropy, $S_{Re}(n)$,  with $S_{EE}=\lim_{n\to 1} S_{Re}(n)$. By using the partial wave expansions of the physical observables  as the total, elastic and inelastic cross sections ($\sigma_{tot}$,  $\sigma_{el}$, and $\sigma_{inel}$) as well as the differential elastic cross section, $d\sigma_{el}/dt$, the entropy is given by,
\begin{eqnarray}
S_{EE}& = & -\lim_{n\to 1} \frac{\partial }{\partial n}\tr_A (\hat{\rho}_A)^n  = -\ln \Omega,\\
\Omega & = & 1- \left( \frac{\sigma_{el}-\frac{4}{f_V}\left. \frac{d\sigma_{el}}{dt}\right|_{t=0}}{\pi f_V - \sigma_{inel}}\right).
\end{eqnarray}
In expression above, $f_V=V/k^2$ with $V:=\sum_{\ell} (2\ell +1)$ being the full phase space volume. Such a volume is formally divergent as the full Hilbert space spans over all partial waves up to $\ell \rightarrow \infty$. In Ref.  \cite{Peschanski:2019yah}  the identification of the physical origin of this divergence and  its further regularization is carefully treated. Finite and regulated  expressions for the entanglement entropy are then applied to $pp$ and $p\bar{p}$ collisions at high energies. Extraction of entropy is performed using the diffractive peak approximation and three different regularization methods. One of them disregards the non-interacting states and an ideal volume regularization is constructed. At the LHC energies the values reach above unity. For instance, at $\sqrt{s}=13$ TeV the entropy for $pp$ collisions is $S_{EE}\approx 1.0370\pm 0.1749$ \cite{Peschanski:2019yah}.

The main goal of this paper is to extract the entanglement entropy using the ideal regularization proposed in \cite{Peschanski:2019yah} (given by Eq. (\ref{SEE-VR}) in what follows) and  the systematic and model-independent method for determining the differential cross section provided by the L\'evy imaging method \cite{Csorgo:2018uyp,Csorgo:2019egs,Csorgo:2020rlb}.  The hadron femtoscopy provided by the L\'evy expansion allows for the reconstruction of the elastic $\bar{p}p$ and  $pp$ scattering amplitudes at low and high energies.  This means that the entropy will be determined in an independent way and its asymptotic limit can be described by absorptive and reflective scattering modes constrained by unitary. For instance, the black disk limit predicted in context of the absorptive mode formalism set a bound $S_{EE}\approx 1+\ln(2)$ for the elastic scattering at asymptotic energies.  This paper is organized as follows. In next section, we start by briefly reviewing the calculation of the entanglement entropy in elastic scattering using $S$-matrix approach and the partial wave expansion (subsection \ref{sec2a}). Also, the phenomenology of L\'evy hadron imaging as applied to the internal structure of the hadrons at collider energies by elastic scattering is reviewed  (subsection \ref{sec2b}).  In Sec. \ref{sec3} the main results are presented  and the uncertainties associated to the formalism and possible future applications are discussed. In Sec. \ref{sec:conc}  we summarize the main results obtained by the analysis.

\section{Theoretical formalism and phenomenology}
\label{sec2}

\subsection{Entanglement entropy in two-body elastic scattering in the $S$-matrix formalism}
\label{sec2a}

First, we shortly review the formalism presented in Refs. \cite{Peschanski:2016hgk,Peschanski:2019yah}, where the entanglement entropy is obtained  for elastic scattering of  two on-shell particles, $A$ and $B$, at high energy regime. It includes also inelastic processes which appear in the overall set of the allowed final states. The reduced density matrix is constructed in  terms of the S-matrix operator projecting the two-body initial state onto the two-body one. The incoming particle 3-momenta are denoted by $(\vec{k},\,\vec{l})$, whereas the outgoing 3-momenta are $(\vec{p},\,\vec{q})$, respectively.  Tracing out the overall density matrix, $\hat{\rho}$,  with respect to the Hilbert space of particle $B$ one obtains,
\begin{eqnarray}
\hat{\rho}_A & = & \rho_0\int \frac{d^3\vp}{2E_{A\vp}}\, 
	\frac{\delta(p-k)\left|\langle \vp,-\vp |{\bf s}|\vk,-\vk \rangle \right|^2}{4k (E_{A\vk} +E_{B\vk})}	
	|\vp\rangle \langle\vp| , \nonumber  \\	
\rho_0^{-1} & = & \delta^{(3)}(0) \int d^3\vp\,
	\frac{\delta(p-k)\left|\langle \vp,-\vp |{\bf s}|\vk,-\vk \rangle \right|^2}{4k(E_{A\vk} +E_{B\vk})} ,  
\label{redden}
\end{eqnarray}
where the following normalization condition is obeyed, $ \tr_A \hat{\rho}_A= \tr_A\tr_B \hat{\rho} = 1 $. This condition leads to the  overall $\delta^{(3)}(0)$ appearing in Eq. (\ref{redden}) and it is the origin of possible divergence in the entropy.  Here, $p=|\vec{p}|$ and $k=|\vec{k}|$ with $\cos \theta =\vec{p}\cdot\vec{k}/(pk)$.

The entanglement entropy is obtained from the reduced matrix in the form $S_{EE} =  -\lim_{n\to 1}\frac{\partial}{\partial n}\tr_A (\hat{\rho}_A)^n$.  After doing the product of the $n$ density operators given at Eq. (\ref{redden}) one obtains $\tr_A(\hat{\rho}_A)^n$ in the following way,
\begin{eqnarray}
\tr_A (\hat{\rho}_A)^n = \int d^3\vp\, \delta^{(3)}(0)
\left(\rho_0\delta(p-k) 
\frac{\left|\langle \vp,-\vp |{\bf s}|\vk,-\vk \rangle 
\right|^2}{4k (E_{A\vk} +E_{B\vk})}
 \right)^n, \nonumber
\label{tracen}
\end{eqnarray}	
where the extra $\delta^{(3)}$  arises from performing the trace over the 3-momentum of the  $A$ particle.  Also, one has the definition $\langle \vp,\vq|S|\vk,\vl\rangle \equiv \delta^{(4)}(P_{p+q} 
-P_{k+l})\, \langle \vp,\vq|{\bf s}|\vk,\vl\rangle$ with the notation $P$ for the center-of-mass 4-vector.  

By making use of the partial wave expansion of the reduced S-matrix element and partial wave expansion of the scattering amplitude \cite{Peschanski:2016hgk,Peschanski:2019yah},
\begin{eqnarray}
\langle \vp,-\vp |{\bf s}|\vk,-\vk \rangle
 &= & \frac{E_{A\vk} +E_{B\vk}}{(\pi k/2)}\left(\delta(1-\cos\theta) + \frac{i {\cal A}}{16\pi}\right),\nonumber \\
{\cal A}(s,t) &=& {16\pi} \sum_{\ell=0}^\infty (2\ell +1) \tau_\ell P_\ell (\cos \theta) ,
 \label{eq:ampli}
\end{eqnarray}
the quantity $\tr_A (\hat{\rho}_A)^n $ can be further computed. In last equation above, $s_\ell =1+2i\tau_\ell$ refers to 2-body S-matrix $\ell^{\rm th}$ partial wave. It can be defined a full phase-space volume, $V \equiv 2\delta(0) = \sum_{\ell=0}^\infty (2\ell +1)$, which is related to the  $\delta^{(3)}$-functions in the form $V=4\pi k^2 \delta^{(3)}(0)/\delta(0)$.

After integration over the 3-momentum and writing Eq. (\ref{tracen}) in terms of the scattering angle $\theta$ and factorizing out the remaining constant factors one obtains,
\begin{eqnarray}
\tr_A(\rho_A)^n &= & \left({V \over 2}\right)^{1-n}\int_{-1}^1 d\cos\theta\, [{\cal P}(\theta)]^n \,, \label{eq:trrhon} \\
{\cal P} (\theta) &= & \delta(1-\cos\theta)\,\left(1 -{2 \sum_\ell(2\ell+1) |\tau_\ell|^2 \over V/2 -  \sum_\ell(2\ell+1)f_\ell}\right) \nonumber \\
&+ & {\left|\sum_\ell(2\ell +1)\tau_\ell P_\ell(\cos\theta)\right|^2 \over V/2 -\sum_\ell(2\ell+1)f_\ell} \,,
\end{eqnarray}
where $f_\ell$ are the partial wave components of the inelastic cross section related to the elastic ones $\tau_\ell$ through the unitarity relation, $f_\ell= 2\bigl(\Im\tau_\ell- |\tau_\ell|^2 \bigr)$. The next step is rewriting ${\cal P} (\theta) $ as a function of the physical observables, $\sigma_{tot},\,\sigma_{el},\, \sigma_{inel}$ and $d\sigma_{el}/dt=|{\cal A}|^2/(256\pi k^4)$, which are usually described in terms of partial wave  components $\tau_\ell$ and $f_\ell$. Namely,
\begin{eqnarray}
{\cal P}(\theta)& = & \delta(1-\cos\theta)\cdot\left(1-{\sigma_{el} \over {\pi V / k^2} - \sigma_{inel}}\right)\nonumber \\
&+ & {2k^2 \over \sigma_{el}}{d\sigma_{el} \over dt}
\cdot\left({\sigma_{el} \over {\pi V / k^2} - \sigma_{inel}}\right),
\label{Pobs}
\end{eqnarray}
with the Mandelstam variable $t=2k^2(\cos \theta -1) $ being the momentum transfer squared.

Finally, the entanglement entropy $S_{EE}$ is properly computed as, 
\begin{eqnarray}
S_{ EE} & = & - \lim_{n \to 1} \frac{\partial}{\partial n} \tr_A (\hat{\rho}_A)^n,\\
& = & \ln {V \over 2} -\int_{-1}^1 d\cos\theta\, {\cal P}(\theta) \ln {\cal P}(\theta) \,.
\label{SEEgeneral}
\end{eqnarray}

In Refs. \cite{Peschanski:2016hgk,Peschanski:2019yah} the authors identified divergences appearing in the calculation of $S_{EE}$ above coming from the divergent full phase-space volume, $V$, as discussed before.  This divergence is interpreted as due to the infinite number of non-interacting 2-body states included for the summation of final states in the derivation. Therefore,  a suitable regularization is need and three options have been suggested in \cite{,Peschanski:2019yah}: (i) volume-regularization, (ii) cut-off (step function) regularization and (iii) cut-off (Gaussian function) regularization. The key feature is that at a given energy the first term in Eq. (\ref{Pobs}) arises from the part of the two-body Hilbert space of the final states which does not correspond to the interacting ones.  A natural way to get rid of those non-interacting modes is regularizing the phase-space volume in order the first term of ${\cal P} (\theta)$ to vanish. Namely, it is defined in such way that $\sigma_{el}/[(\pi\tilde{V}/k^2)-\sigma_{inel}]= 1$. Using the fact that $\sigma_{tot}=\sigma_{el}+\sigma_{inel}$ one gets $\tilde{V}=k^2\sigma_{tot}/\pi$ and accordingly, $\tilde{{\cal P}}(\theta) = \frac{2k^2}{\sigma_{el}}\frac{d\sigma_{el}}{dt}$. This is considered the volume-regularization hypothesis and the volume-regularized entanglement entropy is given by, 
\begin{eqnarray}
 S_{EE} = -\int_0^{\infty} d|t|\, \frac{1}{\sigma_{el}} \frac{d\sigma_{el}}{dt} \ln \biggl(\frac{4\pi}{\sigma_{tot} \sigma_{el}} \frac{d\sigma_{el}}{dt}\biggr). 
 \label{SEE-VR}
\end{eqnarray}
 which depends only on measurable observables. 
 
 In \cite{,Peschanski:2019yah} an estimate of Eq. (\ref{SEE-VR}) was obtained assuming the diffraction peak approximation for hadron-hadron scattering at high energies. In this case, the differential elastic cross section and the elastic cross section are given by,
\begin{eqnarray}
\frac{d\sigma_{el}}{dt} & = & \frac{\sigma_{tot}^2}{16\pi} e^{-B_{el}|t|}, \\
\sigma_{el} & = & \int_0^{\infty}d|t|\,\frac{d\sigma_{el}}{dt} = \frac{\sigma_{tot}^2}{16\pi B_{el}},
\end{eqnarray}
 where $B_{el}(\sqrt{s})$ is the elastic slope parameter which can be written as $B_{el}=\sigma_{tot}^2/(16\pi \sigma_{el})$. 
 
In the diffraction peak approximation, the entanglement entropy in Eq. (\ref{SEE-VR}) therefore becomes 
\begin{eqnarray}
\tilde{S}_{ EE} = 1 +2\ln (2) + \ln \left( \frac{\sigma_{el}}{\sigma_{tot}} \right),
\label{SEE-VRDiff}
\end{eqnarray}
which could be bounded by the black disk limit,  $\sigma_{el}/\sigma_{tot} \rightarrow 1/2$, at asymptotic energies. That is, $\tilde{S}_{EE}(\sqrt{s}\rightarrow \infty)=1+\ln (2)\approx 1.693$.

In order to implement regularization  using an explicit  cut-off, in \cite{,Peschanski:2019yah} the scattering amplitude ${\cal A}$ was rewritten in the impact-parameter representation as,
\begin{eqnarray}
a(s,b)&=&\frac{1}{2\pi}\int d^2q \,e^{-i \,\vec{q}\cdot \vec{b}} f(s,t),\\
f(s,t)&=&\frac{1}{2\pi}\int d^2b \,e^{i \,\vec{q}\cdot \vec{b}} a(s,b),
\end{eqnarray}
where we denote $f(s,t)={\cal A}/(16\pi k^2)$ and thus $\sigma_{tot}=2\int d^2b \,\Im a(s,b)$ and $\sigma_{el}=\int d^2b |a(s,b)|^2$ (with $t=-\vec{q}^2$).

The following prescription is used to approximately obtain the physical Hilbert space. Identifying that $bk\sim \ell$,  the large impact parameter region does not contribute to the scattering amplitude $a (s,b)$ (i.e., the large $\ell$ contribution to partial wave components of the elastic cross section $\tau_{\ell}$). The  regularization procedure is done through the truncation of the large impact parameter modes by introducing a cut-off function $C(b)$ which vanishes at $b \rightarrow \infty$. In this way, the regulated quantities become \cite{Peschanski:2019yah}:
\begin{eqnarray}
\hat{\sigma}_{tot}  & = & 2\int_0^{\infty} d^2b \,C^2(b)\Im a(s,b), \label{tothat}\\
\hat{\sigma}_{el}&= &\int_0^{\infty} d^2b\,C^2(b)|a(s,b)|^2, \label{elhat}\\
\frac{d\hat{\sigma}_{el}}{dt} & = & \frac{1}{4\pi}\left| \int_0^{\infty} d^2b \,e^{i \,\vec{q}\cdot \vec{b}} C(b)\,a(s,b)  \right|^2.  \label{dsdthat}
\end{eqnarray}

Accordingly, the volume of the regularized Hilbert space is given by $\tilde{V}\approx \hat{V}=k^2\hat{\sigma}_{tot}/\pi$ and as a consequence, $\tilde{{\cal P}}(\theta) = \frac{2k^2}{\hat{\sigma}_{el}}\frac{d\hat{\sigma}_{el}}{dt}$. The simplest choices for the function $C(b)$ are the step-function and the Gaussian one. Namely,
\begin{eqnarray}
C(b) & = & 
\begin{cases}
1 & (b \leq 2\Lambda), \\ 
0 & (b > 2\Lambda).
\end{cases}\quad (\mathrm{Step-function}), \label{step} \\
C(b) & = & \exp{\left(-{\small {1 \over 2}}\cdot{b^2 \over 4\Lambda^2}\right)} \quad (\mathrm{Gaussian}). \label{Gaus}
\end{eqnarray}

With regard to the cut-off approximation the entanglement entropy, Eq. (\ref{SEE-VR}), is rewritten as,
\begin{eqnarray}
 \hat{S}_{EE} = -\int_0^{\infty} d|t|\, \frac{1}{\hat{\sigma}_{el}} \frac{d\hat{\sigma}_{el}}{dt} \ln \left(\frac{4\pi}{\hat{\sigma}_{tot} \hat{\sigma}_{el}} \frac{d\hat{\sigma}_{el}}{dt}\right). 
 \label{SEE-CUTOFF}
\end{eqnarray}

Both cutoffs presented above regularize the infinite volume of the Hilbert space because $\ell$ now has an upper bound defined by $\ell_{max}\equiv 2\Lambda k$ and then  $\hat{V}= k^2\hat{\sigma}_{tot}/\pi = 2k^2\int_0^{\infty}\frac{d^2b}{2\pi}C^2(b)=4k^2\Lambda^2$.  Therefore, the condition that determines the cutoff  is $\Lambda^2= \hat{\sigma}_{tot}/4\pi$. For instance, in the forward peak approximation and the Gaussian-function regularization the entanglement entropy is given by \cite{Peschanski:2019yah},
\begin{eqnarray}
 \hat{S}_{EE}^{\mathrm{Gaus}} & = & 1-\frac{4\pi B_{el}\left(1+ \frac{B_{el}}{2\Lambda^2}\right)}{\sigma_{tot}\left(1+ \frac{B_{el}}{2\Lambda^2}\right)} ,\\
 \mathrm{with}\,\,\Lambda & = & \left(\frac{\sigma_{tot}}{4\pi}-\frac{B_{el}}{2}  \right)^{1/2}.
\end{eqnarray}

In Table \ref{table:1} we present the results of calculations done in Ref. \cite{Peschanski:2019yah} using the 3 regularization prescriptions (originally, for $\sqrt{s}=1.8,\,7,\,8$ and $13$ TeV). The measured values of total and elastic cross sections are also presented \cite{PDG2020,Antchev:2017dia}. We also added the predictions for RHIC energy recently measured, $0.2$ TeV \cite{Adam:2020ozo}, and the  LHC data at 2.76 TeV \cite{Antchev:2018rec,Nemes:2017gut}.  Our main goal here is to compute the entanglement entropy with ideal regularization, Eq. (\ref{SEE-VR}), without making use of any assumption about the $t$ dependence (diffraction peak) of the differential elastic cross section and/or any cut-off on impact parameter. To do so, we will employ the model-independent L\'evy imaging method which allow to reconstruct the elastic $pp$ and $p\bar{p}$  scattering amplitudes at both  low and high energies. In what follows, the L\'evy expansion is quickly reviewed focusing on the description of observables needed for computing $S_{EE}$.

\begin{table*}[t]
\caption{The entanglement entropy determined by the model-independent  L\'evy imaging method compared to the diffraction peak approximation presented in Ref. \cite{Peschanski:2019yah}. We present for sake of completeness the results for the three different regularizations schemes (volume regularization and Step/Gaussian function cutoffs). Predictions for 0.2 TeV (RHIC) and 2.76 TeV (LHC) not appearing originally in  \cite{Peschanski:2019yah} are computed.}
\begin{tabular}{|c|c|c|c|c|c|}
\hline
 $\sqrt{s_{pp}}$ (TeV) & L\'evy imaging& Volume-regularization & Exp. data $[\sigma_{tot}, \sigma_{el}] (mb)$ & Step-function & Gaussian-function \\
 \hline
 13.00 & 1.126  &  1.114 & $[110.6 \pm  3.4,\,  31.0 \pm 1.7]$  &  1.212 & 0.8621  \\
 8.00 & --  & 1.063  &  $[101.7 \pm  2.9,\, 27.1 \pm 1.4]$   & 1.197 & 0.7965  \\
 7.00 & 1.020 &  1.031& $[98.0  \pm  2.5,\,  25.1 \pm 1.1]$ & 1.192 & 0.7539 \\
  2.76& -- & 1.029  & $[ 84.7  \pm  3.3,\,  21.8 \pm 1.4]$ & 1.144 & 0.7509 \\
  1.80& 0.953 & 0.918 &  $[72.10 \pm 3.3,\, 16.6  \pm 1.6]$ & 1.193 & 0.6009 \\
  0.20& --&  0.769  &  $[54.67\pm 1.89,\,10.85\pm 0.64]$ & 1.103 & 0.3909 \\
 \hline
 \end{tabular}
 \label{table:1}
 \end{table*}

\subsection{ Model-independent femtoscopic Lévy imaging for elastic scattering }
\label{sec2b}

The  L\'evy series is a generalization of the L\'evy expansion methods proposed to analyze nearly L\'evy stable source distributions in the field of particle femtoscopy \cite{Csorgo:2018uyp,Csorgo:2019egs,Csorgo:2020rlb}. Here, we are interested in the momentum transfer $t$-distribution in hadron-hadron elastic collisions.  It provides  a systematic and model-independent method to characterize the variations from the approximate shape of these distributions  by making use of a dimensionless variable, $z \equiv R^2|t| \geq 0$,  and a complete orthonormal set of polynomials that are orthogonal with respect to the weight function $\omega (z)=\exp(-z^{\alpha})$.  The quantity $R$ denotes the L\'evy scale parameter. We follow closely the recent analysis of differential elastic $pp/p\bar{p}$ scattering cross-sections done in Refs.~\cite{Csorgo:2018uyp}. A clear advantage of the L\'evy method for proton imaging is supplying the  inelasticity profile of the proton as a function of energy and impact parameter. In momentum  $t$-representation, the elastic differential cross-section is related to the modulus of the complex-valued elastic amplitude $T_{el}$. The latter is expressed as an orthonormal series expansion  in terms of the L\'evy  polynomials~\cite{Csorgo:2018uyp},
\begin{eqnarray}
\frac{d\sigma_{el}}{dt} & = & \frac{1}{4\pi}|T_{el}(s,t)|^2,\\
T_{el}(s,t) & = &  i\sqrt{4\pi A}\, e^{- \frac{z^{\alpha}}{2}}\left( 1+\sum_{i = 1}^\infty c_i l_i (z|\alpha) \right),
\label{LevyTel}
\end{eqnarray}
where $c_i = a_i + i b_i$ are the complex expansion coefficients ($a_i$ and $b_i$ being the real and the imaginary parts of $c_i$, respectively). The quantities $l_j(z|\alpha)$ are the normalized L\'evy polynomial 
of order $j$, which are given by, 
\begin{eqnarray}
        l_j(z\, |\, \alpha) & = & \frac{ L_j(z\, |\, \alpha)}{\sqrt{ D_{j}(\alpha)}\,\sqrt{ D_{j+1}(\alpha)}},\, \mathrm{for}\,\, j\ge 0 \, .
    \label{ljs}
\end{eqnarray}
constructed in terms of the the unnormalized L\'evy polynomials $L_i(z\,|\,\alpha)$ (where one has $L_0(z\,|\,\alpha) = 1$),
\begin{eqnarray}
L_1(z\,|\,\alpha)  &=& 
       \det\left(\begin{array}{cc}
     \mu_{0}^{\alpha} & \mu_{1}^{\alpha}  \\ 
     1 & z \end{array} \right) \,, \\
 L_2(z\,|\,\alpha)  &=& 
       \det\left(\begin{array}{ccc}
     \mu_{0}^{\alpha} & \mu_{1}^{\alpha} & \mu_{2}^{\alpha} \\ 
     \mu_{1}^{\alpha} & \mu_{2}^{\alpha} & \mu_{3}^{\alpha}  \\ 
     1 & z & z^2 \end{array} \right), \\    
      L_m(z\,|\,\alpha)  &=& 
       \det\left(\begin{array}{ccc}
     \mu_{0}^{\alpha} & \cdots & \mu_{m}^{\alpha} \\ 
    \vdots & \ddots & \vdots \\
     1 & \cdots & z^m \end{array} \right), 
\end{eqnarray}
and the Gram-determinants, $ D_j(\alpha)$, are defined by
\begin{eqnarray}
D_1(\alpha)  &=&  \mu_{0}^{\alpha} \,, \quad
D_2(\alpha)   = 
     \det\left(\begin{array}{cc}
     \mu_{0}^{\alpha} & \mu_{1}^{\alpha}  \\ 
     \mu_{1}^{\alpha} & \mu_{2}^{\alpha} 
     \end{array} \right)\,,  \\
     D_m(\alpha)   & = & 
     \det\left(\begin{array}{ccc}
     \mu_{0}^{\alpha} &  \cdots & \mu_{m-1}^{\alpha}  \\ 
      \vdots & \ddots & \vdots \\
     \mu_{m-1}^{\alpha} &  \cdots&  \mu_{2m-2}^{\alpha} 
     \end{array} \right)\,,  \\
    \mu_{n}^{\alpha} &=&  \frac{1}{\alpha}\,\Gamma\left( \frac{n+1}{\alpha}\right) \,.
\end{eqnarray}
 where $D_0(\alpha) \equiv  1$ and $\Gamma(x) $ is the Gamma function.
 
The total cross section, $\sigma_{tot}  \equiv   2\Im T_{el}(s,0)$,  and elastic cross-sections are expressed in terms of the quantities defined above, 
\begin{eqnarray}
\sigma_{tot} & = &2\,\sqrt{4\pi A}\,\left(1 + \sum_{i = 1}^\infty a_i l_i (0|\alpha) \right),
    \label{SIGTOT} \\
    \sigma_{el}  & = & \frac{A}{R^2} \, \left[\frac{1}{\alpha}\Gamma\left(\frac{1}{\alpha}\right) +  \sum_{i = 1}^\infty (a_i^2 + b_i^2) \right].
    \label{SIGEL}
\end{eqnarray}
It was demonstrated in Ref. \cite{Csorgo:2018uyp} that the expansion for $T_{el}(s,t)$ converges very fast and a third-order L\'evy series is enough to reproduce the data measured at $\sqrt{s}\leq 1$ TeV with confidence levels corresponding to a statistically suitable description.
In next section, the L\'evy imaging method will be used to compute the entanglement entropy in the ideal regularization scheme at high energies. The low energy data are considered as well. We used the results of the fourth-order L\'evy expansion to the elastic scattering data of proton-proton collisions measured in the ISR energy range ($\sqrt{s}= 23.5, 30.7, 44.7, 52.8$ and $62.5$ GeV). Moreover, for proton-antiproton collisions a second-order expansion is used for energies of $\sqrt{s} = 53$ GeV (ISR) and  $\sqrt{s} = 1960$ GeV (D0, Tevatron) whereas a third-order expansion stands for $\sqrt{s} = 546$ GeV and  $\sqrt{s} = 630$ GeV (UA4). For the LHC energies, a fourth-order expansion to all the differential cross section measurements of elastic $pp$ collisions at 7 and 13 TeV has been taken. The parameters of the expansions, $R,\,\alpha$ and the complex coefficients $c_i$ are available in Appendices A and B of Ref. \cite{Csorgo:2018uyp} and in Refs. \cite{Csorgo:2019egs,Csorgo:2020rlb}. Typically, one has $\alpha \simeq 0.9$ and $R\simeq 0.6-0.7$ fm. In next section the model independent extraction of $S_{EE}$ is compared with those from Ref. \cite{Peschanski:2019yah} and an analysis on its energy dependence is performed using for simplicity the eikonal model.

\begin{figure*}[t]
		\includegraphics[scale=0.4]{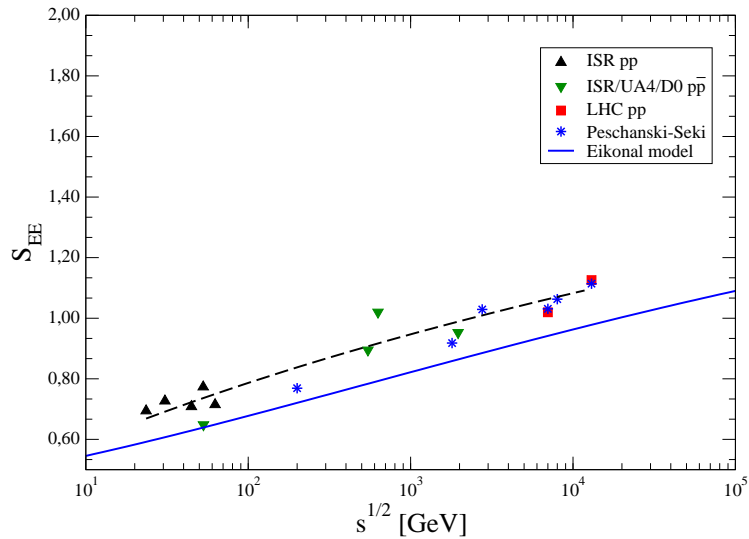} 
	\caption{The entanglement entropy for elastic scattering as a function of center-of-mass collision energy, $\sqrt{s}$. Extraction using L\'evy imaging method is presented at low and high energies and compared to the results from Ref. \cite{Peschanski:2019yah}. The values for LHC, Tevatron and RHIC energies are presented in Table \ref{table:1}. Prediction for  diffraction peak approximation using the one-channel eikonal model is shown (solid line). A fit based on the single Regge pole contribution to the soft Pomeron is also presented (dashed line).}
	\label{fig:1}
\end{figure*}

\section{Results and discussions}
\label{sec3}

In Fig. \ref{fig:1} is shown the extracted entanglement entropy as a function of center of mass energy using the ideal volume regularization scheme, Eq. (\ref{SEE-VR}) using the L\'evy imaging method. The low energy data for $pp$ collisions from ISR are labeled by up triangles, whereas the $p\bar{p}$ collision data from ISR, UA4 and D0 are represented by down triangles. The TOTEM-LHC data at 7 and 13 TeV are presented (squares) together with extracted values for $S_{EE}$ in Ref. \cite{Peschanski:2019yah} (stars) using the same ideal volume regularization. The L\'evy expansion gives somewhat large values of entanglement entropy using the ideal regularization due to the additional contribution at large $t$ which is suppressed in the diffraction peak approximation considered in \cite{Peschanski:2019yah}. However, the deviation is not so high and the small-$t$ approximation can be considered a suitable extraction for $S_{EE}$. For sake of completeness, in Table \ref{table:1} is presented the comparison between the different extraction methods at high energies (LHC and Tevatron) and the values of cross section measurements. We included also the recent results for $\sigma_{tot}$ and $\sigma_{el}$ in $pp$ collisions for RHIC at $\sqrt{s}=200$ GeV.  We verified that the step-function regularization option is numerically time consuming due to the oscillating integrand in Eq. (\ref{dsdthat}).

At low energies, the proton-proton elastic scattering (ISR) presents an entanglement entropy of order 0.7. The proton-antiproton scattering at intermediate and high energies (UA4, Tevatron) provides $S_{EE}\simeq 1$. At the LHC energies the entropy reaches values around 1.2 at 13 TeV. It would be worth obtaining the L\'evy expansion extraction in the intermediate LHC energies of 2.76 and 8 TeV in order to confirm the trend on the behavior as a function of the center of mass energy. As we will see in what follows it is roughly expected that the  entropy in forward peak approximation, $\tilde{S}_{EE} \sim \ln (\sigma_{tot}/\sigma_{el})$, saturates at very high energies. We have also discussed the bound given by the black disk (BD)  limit, $\sigma_{tot}/\sigma_{el}\rightarrow 1/2$,  which corresponds to the maximal absorption within the eikonal unitarization. On the other hand, in the U-matrix formalism the scattering amplitude \cite{Troshin:2007fq} in impact parameter space  may exceed the BD limit with the colliding particles becoming progressively more transparent \cite{Troshin:2007fq,Troshin:2019ivz}, i.e. the gray disc limit. In this unitarization scheme, that ratio reaches its maximal possible value, $\sigma_{tot}/\sigma_{el}\rightarrow 1$, at asymptotic energies. Specifically, this means that when the  amplitude exceeds the BD limit then the scattering becomes driven by anti-shadow contribution \cite{Troshin:2019ivz}.  For the anti-shadow mode the elastic amplitude in impact parameter space increases with decrease of the inelastic channels pieces.

In order to shed light on the energy dependence of the entanglement entropy in high energy elastic collisions we will consider the one-channel eikonal model in impact parameter space (GLM model) \cite{Gotsman:1992ui,Gotsman:1993vd}. The reason is that the ratio $R_{el}(\sqrt{s})=\sigma_{el}/\sigma_{tot}$ can be analytically evaluated.  In the diffraction peak approximation and ideal volume regularization, $\tilde{S}_{EE}=\ln(4eR_{el})$. Using s-channel unitarity and a simplified form for the scattering amplitude in impact parameter representation, i.e. $a(s,b)=i[1-e^{\frac{\Omega(s,b)}{2}}]$, the total, elastic and inelastic cross section can be easily computed. The Opacity function, $\Omega$, is written in terms of a single $t$-channel soft Pomeron ($\pom$) exchange in a factorized way $\Omega (s,b)=g(s)S(s,b)$ with the notation $\nu (s)=\Omega (s,0)$ \cite{Gotsman:1992ui,Gotsman:1993vd}. The quantity $S(s, b)$ is the b-space normalized soft profile function. By using a Gaussian soft profile the Opacity takes the form,
\begin{eqnarray}
\Omega (s,b) = \frac{\sigma_0}{2\pi B_{el}}\left( \frac{s}{s_0} \right)^{\Delta_{\pom}}\exp \left(-\frac{b^2}{2B_{el}}  \right),
\end{eqnarray}
 where $B_{el}=B_0+2\alpha^{\prime}_{\pom}\ln(s/s_0)$ and $B_0$ is the slope of the elastic differential cross section due to the Pomeron exchange at $s = s_0$. It is well known that the ratio $R_{el}$ takes the analytical form,
 \begin{eqnarray}
 R_{el}=\frac{1}{2}\frac{\ln (\nu/4)+\gamma_E-\Gamma(0,\nu)+2\Gamma(0,\nu/2)}{\ln (\nu/2)+\gamma_E+\Gamma (0,\nu/2)}, 
 \label{releik}
 \end{eqnarray}
 where $\gamma_E \approx 0.5772$ is the Euler-Marscheroni constant and $\Gamma(0,x)$ is the incomplete Gamma function.  We took the parameters from Ref. \cite{Gotsman:2005wa}, which does not includes the LHC data.  The fitted values for the Pomeron parameters in the one-channel eikonal  model are $\Delta_{\pom}=0.09$, $\alpha^{\prime}_{\pom}=0.25$ GeV$^{-2}$, $s_0=450$ GeV, $\sigma_0=47.2$ mb and $B_0=10.24$ GeV$^{-2}$ \cite{Gotsman:2005wa}. Here, the main goal is to obtain an analytical expression of $S_{EE}$ as a function of energy. The adjusted parameters can be of course updated using the recent LHC data, which is out of scope of present study (it is known that LHC data on soft scattering are only marginally compatible with the simple soft Pomeron GLM model \cite{Gotsman:2012rq}). For instance, the measured value at the LHC by the TOTEM Collaboration at 13 TeV is $R_{el}=0.281\pm 0.009$  \cite{Antchev:2017dia} and the eikonal model gives  $R_{el}^{eik} = 0.23$. On the other hand, one-channel  eikonal models based on QCD with  nonperturbative
effects included through a QCD effective charge are able to successfully describe the LHC data \cite{Broilo:2019yuo,Broilo:2019xhs}. At high energies, Eq. (\ref{releik}) can be further simplified as $\Gamma (0,\nu\gg 1)\approx \nu^{-1} e^{-\nu}$. For instance, $\nu(\sqrt{s}=13 \,\mathrm{TeV})=3.68$ and a good approximation at very high energies is $R_{el}\approx \frac{1}{2}[\frac{\ln(\nu/2)}{\gamma_E}+1]^{-1}$. This implies an energy behavior for the entropy  like $S_{EE}\sim \ln (\ln s)$.
 
 Here, some words are in order. It is a complex task to single out the energy dependence of the entanglement entropy using the regulated entropy in Eq. (\ref{SEE-VR}) as the final result after $t$-integration is strongly dependent on the specific behavior of the elastic differential cross section at small and large $t$ (we quote Ref. \cite{Pancheri:2016yel} where a comprehensive review is done on the hadron cross sections from lower to the highest energies, including comparison to several models). This task is quite simplified in the diffraction peak approximation and the one-channel eikonal model describes analytically its energy behavior. In this case, the physical parameters driving the s-dependence of the entropy  are the soft Pomeron intercept $\alpha_{\pom}(0)$, with $\Delta_{\pom}=\alpha_{\pom}(0)-1$, and the elastic slope $B_{el}(s)\sim \ln (s)$. Particularly, $S_{EE}\sim \ln (\ln \nu(s))$ with $\nu$ being the Opacity at central impact parameter, $\nu(s)=\Omega (b=0)\sim s^{\Delta_{\pom}}/B_{el}$. Using the traditional Regge phenomenology and taking the single pole contribution to the soft Pomeron we can investigate from which property of observables the energy dependence of $S_{EE}$ comes from. In this picture, the total cross section is given by $\sigma_{tot}=4\pi g_{\pom}s^{\alpha_{\pom}(0)-1}$ with $g_{\pom} =\gamma_{\pom}(0)\mathrm{Im}\eta_{\pom}(0)$. The quantities $\gamma_{\pom}$ and $\eta_{\pom}$ are the residue function at the pole and the signature factor at $t=0$, respectively. Assuming a simple exponential form for the residue function, $\gamma_{\pom}(t) = \gamma_{\pom}(0)\exp (-B_0|t|)$, one has that the elastic differential cross section given by $d\sigma_{el}/dt \sim s^{2(\alpha_{\pom}-1)}\exp (Bt)$. The effective slope of the elastic amplitude for the linear Pomeron trajectory, $\alpha_{\pom}(t)=\alpha_{\pom}(0)+\alpha^{\prime}_{\pom} t$, is given by $B(s)=B_0+2\alpha^{\prime}_{\pom} \ln (s)$. In this approximation is straightforward to obtain $\sigma_{el}\sim s^{2(\alpha_{\pom}(0)-1)}/B(s)$ and putting all together in Eq. (\ref{SEE-VR}) and performing the $t$-integration we find that:
 \begin{eqnarray}
 S_{EE} \sim \ln \left( \frac{4\pi B(s)}{s^{\alpha_{\pom}(0)-1}} \right)\sim  \ln \left( \frac{B_0+2\alpha^{\prime}\ln (s)}{s^{\Delta_{\pom}}} \right).
 \label{sdep}
 \end{eqnarray}
 This means that $s$-dependence of the entropy is influenced mostly by the total cross section and the effective slope. In order to test the reliability of the relation (\ref{sdep}) we performed a three parameter fit to the extracted $S_{EE}$ in the form $S=\ln [a_0(s/s_0)^{-\Delta_{\pom}}(a_1+0.5\ln(s/s_0))]$ with fixed $s_0=100$ GeV$^2$ (in the energy range $23.5\leq \sqrt{s}\leq 13000$ GeV). It is found $\Delta_{\pom}=0.0019\pm 0.0022$, $a_0=0.1775\pm0.0140$ and $a_1=10.1759\pm 0.8466$ and the results is represented by the dashed curve in Fig. \ref{fig:1}. The fit is consistent with a soft Pomeron with intercept $\alpha_{\pom}(0)\approx 1$ and we have checked that by using the standard supercritical Pomeron intercept $\Delta_{\pom}=0.08$  the description is reasonable for pre-LHC energies and underestimates the entropies at LHC energy range. This is due to the unitarity corrections disregarded in the single pole approximation and expected to play a significant role at the LHC. Namely, for the inputs $\sigma_{tot}\sim s^{\Delta_{\pom}}$ and $\sigma_{el}\sim s^{2\Delta_{\pom}}/B(s)$ the power behavior is modified to $\sigma_{tot}, \sigma_{el}\sim \ln^2(s)$ in unitarized models \cite{Gotsman:1992ui,Gotsman:1993vd}. 
 
 The prediction for $\tilde{S}_{EE}(\sqrt{s})=1+2\ln(2)+\ln[R_{el}(\sqrt{s})]$ using the one channel eikonal model (GLM model) is presented in Fig. \ref{fig:1} (solid line). We included also the Reggeon contribution to the Opacity as low energy data are also presented. The eikonal prediction underestimates the data points as the parameters were fitted without including the LHC data as discussed before. The black-disc limit, $R_{el}\rightarrow 1/2$, imposes a limitation on the entanglement entropy for two-body elastic scattering in hadron collisions. Namely, $\tilde{S}_{EE}\rightarrow \ln(2e)\approx 1.693$ at asymptotic energies. Of course, this is the case for the absorptive scattering mode whereas the predicted ratio $R_{el}$ in the reflective scattering mode is somewhat different \cite{Troshin:2007fq}. The elastic scattering would have an absorptive nature in the energy region $\sqrt{s}\lsim 5$  TeV, where elastic and inelastic cross sections  obey the relation $\sigma_{inel}(s) \leq \sigma_{el}(s)$.  Above some energy threshold, $s_r$,  defined as $S(s_r,b=0)=0$  the scattering picture at small $b$ gradually acquires a reflective contribution. In this region, $\sigma_{inel}<\sigma_{tot}-\pi r^2(s)$, where $S(s,r(s))=0$. The value of the ratio $R_{el}(s)$ at $s > s_r$ is correlated with the degree of reflection, while the value of the ratio $R_{inel}(s)=\sigma_{inel}/\sigma_{tot}$  (with $R_{el}+R_{inel}=1$ due to unitarity) is correlated with the degree of the so called hadron hollowness \cite{Troshin:2019ivz}. It is claimed that at the LHC energies of 8 and 13 TeV reflective scattering mode starts to take place and a speed up of the ratio $R_{el}$ is expected. Asymptotically, including the reflective mode the limit $R_{el}(s\rightarrow \infty)\rightarrow 1$ is predicted and the entanglement entropy would have a higher bound, $\tilde{S}_{EE}\rightarrow \ln(4e)\approx 2.386$. We quote Ref. \cite{Fagundes:2015lba} where model-independent analytical parameterizations of the ratio $R_{el}$ as a function of energy are investigated focusing on its asymptotic limit.
 
Concerning nuclear targets one question that arises is what is the value of the entanglement entropy for proton-nucleus elastic scattering. We have no answer by now, but if the ideal regularization expression Eq. (\ref{SEE-VRDiff}) remains the same on $pA$ collisions then some estimate can be done. In Ref. \cite{Bondarenko:2000uv} a generalization of the Glauber-Gribov formalism for $pA$ and $AA$ collisions is proposed which takes into account  the usual rescattering of the fastest partons and the interaction of all partons with the target and the projectile. As an example, at energy $\sqrt{s_{pA}}=2$ TeV and for gold nucleus ($A=197$) one obtains $\sigma_{tot}^{pA}\approx 5.1$b and  $\sigma_{el}^{pA}\approx 1.9$b. This would give $\tilde{S}_{EE}\approx  1.40$,  bearing in mind that the theoretical estimates for nuclear interaction suffer of large uncertainties \cite{Alvioli:2013vk,Kopeliovich:2005us} (in \cite{Goncalves:2019agu} smaller values for total and elastic $pA$ cross section are obtained by using Miettinen- Pumplin model for $pp$ collisions and fluctuation effects in the  nuclear case). 

As a final comment, there have been strong efforts to investigate the relation between the entanglement entropy and the properties of the holographic QCD models \cite{Ali-Akbari:2017vtb,Dudal:2018ztm,Mahapatra:2019uql,Lezgi:2019fqu,Li:2020pgn}. Namely,  in the AdS/QCD correspondence the holographic duality \cite{Kim:2012ey} of entanglement entropy between boundary region $A$ and its complement is the holographic entanglement entropy (HEE), $S_A^h$,   which is obtained by using the Ryu-Takayanagi relation \cite{Ryu:2006bv,Ryu:2006ef}. The latter is a generalization of the proportionality of black hole entropy to the area of its event horizon. The HEE is equivalent to the area of minimal 3-dimensional surface in the bulk which is homologous to $A$, the so called Ryu-Takayanagi surface $\gamma_A$, over a constant value equals to $4G_N$.  Explicitly, $S_A^h\equiv \frac{2\pi}{\kappa^2}\mathrm{Area}(\gamma_A)$ with $\kappa^2=8\pi G_N$ being the gravitational constant.  The usual procedure for evaluating the HEE is to set a region and getting  a finite area of the minimal surface properly UV/IR regulated. The entanglement entropy determination in soft processes presented here is eminently driven by non-perturbative aspects of QCD and holographic methods can shed some light on the problem. Along these line, recent studies put forward the calculation of the total and elastic cross sections at high energies by using the bottom-up AdS/QCD models in the five-dimensional AdS space \cite{Watanabe:2018owy,Xie:2019soz,Watanabe:2019cvw}. Moreover, sophisticated approaches taking into account completely anisotropic holographic models containing different spatial scale factors have been proposed \cite{Arefeva:2020uec,Arefeva:2019dvl}. In these models describing aspects of heavy ion collisions  there is a relation between anisotropy of the background and anisotropy of the heavy ions geometry. It was found that the holographic entanglement entropy and its density have important fluctuations near the black hole phase transition line in chemical potential-temperature plane for all values of the anisotropy parameter. In addition, it was demonstrated that the HEE entanglement entropy of the colliding ions  is further related  with the multiplicity of particles produced \cite{Arefeva:2019dvl}.

\section{Summary}
\label{sec:conc}

We have studied the entanglement entropy for high energy elastic scattering in $pp$ and $\bar{p}p$ collisions, which is theoretically obtained by the $S$-matrix formalism and partial wave expansion of physical observables. It was extended the seminal analysis  done in Ref.\cite{Peschanski:2019yah} which used the diffraction peak approximation and a model independent extraction has been performed using the L\'evy imaging method.   The ideal volume regularization is considered as it is involves only the measured quantities in soft region like the elastic and total cross section and the elastic differential cross section as well. The femtoscopy of hadrons allowed by this expansion method opens the possibility for a systematic extraction of entanglement between the final state hadrons and at high energies $S_{EE}\sim 1$. We discuss the theoretical bound for the entanglement entropy coming from black disc or gray disc limits related to the inclusion of absorptive and reflective scattering modes. We verified that at high energies the entropy for elastic scattering behaves parametrically like $S\sim 1+\ln(2)-\ln (\ln(s))$ and saturating at asymptotic energies. To gain some insight about the energy dependence of the entropy we make use of the simple one-channel eikonal model and qualitative and quantitative analysis has been done. It was found that the energy dependence of the entropy is influenced mostly by the total cross section and the effective slope, $B$, where $ S_{EE} \sim \ln \left( \frac{4\pi B(s)}{s^{\alpha_{\pom}(0)-1}} \right)$. It is possible to describe the extracted entanglement entropy using the single pole contribution to soft Pomeron with an intercept close to unit or by using the GLM eikonal model. The present study is somewhat complementary to our investigation in Ref. \cite{Ramos:2020kaj}, where the entanglement of gluons in DIS off proton and nuclei was addressed.

In summary, the present study carefully investigates the entanglement entropy in soft scattering processes using systematic and  model independent tools which are helpful to single out the main aspects of the entangled final state. The knowledge about entanglement described in a non-perturbative sector of QCD is deeply related to the holographic entanglement entropy in the context of holographic models of QCD  \cite{Ali-Akbari:2017vtb,Dudal:2018ztm,Mahapatra:2019uql,Lezgi:2019fqu,Li:2020pgn}. These models based on AdS/QCD duality are shown to be promising as they are able to describe the main observable as total and elastic cross sections \cite{Watanabe:2018owy,Xie:2019soz} and heavy-ions observables as well  \cite{Arefeva:2020uec,Arefeva:2019dvl}. In former models the proton gravitational form factor  can be obtained from the $p\pom (\mathrm{graviton})p$ three-point function.

\begin{acknowledgments}
We are thankful to Irina Ya. Aref’eva (Steklov Math. Inst., Moscow) for bringing to our attention the concepts of the HEE for heavy ion collisions in anisotropic background with confinement-deconfinement phase transition. This work was  partially financed by the Brazilian funding agencies CNPq and CAPES.
\end{acknowledgments}

\end{document}